\documentclass[a4paper,10pt,twoside]{cpc-hepnp}

\usepackage{multicol}
\usepackage{graphicx}
\usepackage{epstopdf}
\usepackage{booktabs}
\usepackage{amssymb,bm,mathrsfs,bbm,amscd}
\usepackage[tbtags]{amsmath}
\usepackage{lastpage}
\usepackage{xcolor}

\def\pdellx'{\frac{\partial}{\partial x'}}
\def\pdellw'{\frac{\partial}{\partial w'}}

\newcommand{\be}{\begin{equation}}
\newcommand{\ee}{\end{equation}}
\def\bed{\begin{displaymath}}
\def\eed{\end{displaymath}}
\def\bea{\begin{eqnarray}}
\def\eea{\end{eqncrray}}
\def\[{$$}
\def\]{$$}

\begin{document}

\fancyhead[c]{\small Chinese Physics C~~~Vol. xx, No. x (201x) xxxxxx}
\fancyfoot[C]{\small 010201-\thepage}

\footnotetext[0]{Received }

\title{The weak cosmic censorship conjecture and thermodynamics in the quintessence AdS black hole  under charge particle absorption}
\author{%
Ke-Jian He(ºÎ¿Â½¡)$^{1}$, Xin-Yun Hu(ºúöÎÔÈ)$^{2,1)}$\email{huxinyun@126.com}, Xiao-Xiong Zeng(ÔøÏþÐÛ)$^{3,4}$ %
}
\maketitle

\address{%
$^1$ Physics and Space College, China West Normal University, Nanchong 637000, China
}

\address{%
$^2$College of Economic and Management, Chongqing Jiaotong University, Chongqing 400074, China}

\address{%
$^3$Department of Mechanics, Chongqing Jiaotong University, Chongqing 400074, China}

\address{%
$^4$State Key Laboratory of Mountain Bridge and Tunnel Engineering, Chongqing Jiaotong University,\\ Chongqing 400074, China}

\begin{abstract}
   Considering the cosmological constant  as the pressure, we mainly study the  laws of thermodynamics and weak cosmic censorship conjecture in the  Reissner-Nordstr\"{o}m-AdS  black hole surrounded by quintessence dark energy under charged  particle absorbtion. The  first law of thermodynamics is found to be valid as a particle is absorbed by the black hole. The second law however is found to be violated for the extremal  and near-extremal black holes since the entropy of these black hole decrease. Moreover, we find that the extremal  black hole do not change it configuration in the extended phase space, implying that the  weak cosmic censorship conjecture is valid. Remarkably, the near-extremal  black hole can be overcharged beyond the extremal condition under charged particle absorption. That is, the cosmic censorship conjecture could be violated for the near-extremal  black hole in the extended phase space.  To make a comparison, we also discuss the   first law, second law  as well as  the weak cosmic censorship conjecture  in the normal phase space, and find that all of them are valid in this case.
\end{abstract}

\footnotetext[0]{\hspace*{-3mm}\raisebox{0.3ex}{$\scriptstyle\copyright$}2019
Chinese Physical Society and the Institute of High Energy Physics
of the Chinese Academy of Sciences and the Institute
of Modern Physics of the Chinese Academy of Sciences and IOP Publishing Ltd}%

\section{Introduction}

Black holes are very  important not only in astronomy but also in  gravitation theory. For a black hole, there is a  geometric singularity  located at its center. If the event horizon of the black hole is unstable, the singularity of the black hole will be exposed. The  exposed   singularity will send uncertain information, leading to the broken of the causal relation between space and time. To avoid this situation, Penrose conjectured that the singularity should be wrapped in the center of  the event horizon of the black hole, which is the so-called weak cosmic censorship conjecture \cite{ref1, ref2}. This conjecture also illuminates that the  gravitational collapse of dust will eventually form a black hole with singularity \cite{ref2}, which was proved by Penrose and Hawking \cite{ref3}.

  Since there is no general method to prove the weak cosmic censorship conjecture in black holes, we need to check its   validity  in various spacetimes.
 One interesting method is to check whether the final states of  black holes  are still black holes after the absorption. In the four-dimensional space time, Wald proved that particles making an extremal Kerr-Newman black hole overcharge or overspin  would not be absorbed through Gedanken experiment \cite{ref4}. This result is also applicable to scalar field \cite{ref5,ref6}. Later, Hubeny found that for a near-extremal  Reissner-Nordstr\"{o}m  black hole, the weak cosmic censorship conjecture was violated for the charge can exceed its extremum boundary \cite{ref7}.  This behavior was also observed in the near-extremal  Kerr and Kerr-Newman black hole \cite{ref8,ref9}. However, as the  self-force effect was considered in the Kerr-Newman black hole, the weak cosmic censorship conjecture was found to be still valid for   the event horizon  was stable \cite{ref10,ref11,ref12}.  In addition, the near-extremal Reissner-Nordstr\"{o}m  black hole are also  valid for this conjecture with consideration of  the back-reaction effect \cite{ref13,ref14}. Many studies have been made on the validity of the weak cosmic censorship conjecture under different black hole backgrounds \cite{ref15,ref16,ref17,ref18,ref19,ref20,ref21,ref22,ref23,ref24,ref25,ref26,ref27,ref28,ref29,ref30,ref31,ref32,ref33,ref34,ref35,a1,a2,a3,Zeng:2019aao,Zeng:2019baw}, such as black holes in Einstein's gravity theory, modified gravitational theory, black holes with low or high dimensions, and black holes with electric charge .

The cosmological constant is fixed initially. But with some later studies, the cosmological constant was considered  as a variable quantity \cite{ref39,ref40} and the extended phase space is constructed, where the pressure of the black hole is related to the cosmological constant \cite{ref41,ref42},  and its thermal conjugate is the thermodynamic volume of the black hole \cite{ref43,ref44}.
In this case,  the mass of the black hole  corresponds to the enthalpy, not  the internal energy  of the black hole thermodynamic system \cite{ref45,ref46}.

In the extended phase space, the first law of thermodynamics has been investigated extensively. However,  there is little work to investigate the second law and the weak cosmic censorship conjecture. The validity of the first law dose not mean the second law as well as the   weak cosmic censorship conjecture are valid, it thus important and necessary to study  the second law  and the   weak cosmic censorship in the extended phase space.
Recently, Ref. \cite{Gwak:2017kkt}  investigated the first law, second law, as well as the weak cosmic censorship conjecture  in the  Reissner Nordstr\"{o}m-AdS black holes. It was found that the first law and the weak cosmic censorship conjecture were valid, the second law   was violated  for the extremal and  near-extremal black holes. In this study, we will extend the idea in Ref. \cite{Gwak:2017kkt} to the black holes with
quintessence dark energy. We  want to explore whether the second law is valid. As a result, we find that the second law is  violated too  for the extremal and near-extremal black holes. In other hand, it a note that we investigate the change of the minimum value of the metric function  at the higher order in the process of particle absorption. This is different from their research. In Ref. \cite{Gwak:2017kkt}, they think that $\mathcal O(\epsilon^2)$ is the high order terms of $\epsilon$, so it can be considered to zero. However, we can not easily ignore the value of the high order term due to  we can not determine the value  magnitude  of the $\delta_\epsilon$ and $\mathcal O(\epsilon^2)$. Hence, we have extended the  investigate of the changes of the minimum value of the function to the higher order. Moreover, our results show that the weak cosmic censorship conjecture could be violated for the near-extremal black hole in the extended phase space. This result has not appeared in previous studies.

The remainder of this article is organized as follows. In section 2, we briefly review the thermodynamics of the   Reissner-Nordstr\"{o}m-AdS   black hole surrounded by quintessence dark energy.  In section 3, we establish the first law of thermodynamics   in the extended phase space firstly, and then we discuss the second law and weak cosmic censorship conjecture in this framework. In section 4, we investigate  the first law, second law as well as the weak cosmic censorship conjecture in the normal phase space.  Section 5 is devoted to our conclusions.

\section{Review of the Reissner-Nordstr\"{o}m-AdS black hole  surrounded by quintessence dark energy}

The metric of the spherically symmetric charged-AdS black hole surrounded by quintessence dark energy  can be written as \cite{ref48}
\begin{equation}
ds^{2}=-f(r)dt^{2}+f^{-1}(r)dr^{2}+r^{2}(d\theta^2+\sin^2\theta d\phi^2),\label{metric1}
\end{equation}
with
\begin{equation}
f(r)=1-\frac{a}{r^{3 \omega+1}}+\frac{r^2}{l^2}-\frac{2 M}{r}+\frac{Q^2}{r^2},\label{metric}
\end{equation}
the electric potential of the black hole is
\begin{align}
A_\mu =(-\frac{Q}{r}, 0, 0, 0). \label{eq2.3}
\end{align}
In the above equation, $M$ and $Q$ are the ADM mass and charge of the black hole, $l$ is the radius of the $AdS$ spacetime which is related to the cosmological constant. And $a$ is the normalization factor closely related to quintessence density and should be greater than zero. There are several circumstances for the value of $\omega$,  for $-1 <\omega<-1/3$, it is quintessence dark energy, and for $\omega<-1$, it is phantom dark energy. The values of state parameter $\omega$ affect the structure of spacetime. When $\omega=-1$, it  affects the  AdS radius, while when $\omega=-1/3$, it affects the curvature $\kappa$ of space-time \cite{ref49}.

At the event  horizon $r_h$, the Hawking-temperature, Bekenstein-Hawking entropy and electric potential can be expressed as
\begin{align}
T_h=\frac{f'(r_h)}{4 \pi}=\frac{ r_h\left(3a \text{$\omega $r}_h{}^{-3 \omega }+3r_h{}^3/l^2+r_h\right)-Q^2}{4\text{$\pi $r}_h{}^{3 }},  \label{eq2.4}
\end{align}
\begin{align}
S_{h}=\pi r_h^2,  \label{eq2.5}
\end{align}
\begin{align}
\Phi _h=-A_t(r_h)=\frac{ Q}{r_h}.  \label{eq2.6}
\end{align}
 In  the extended thermodynamic phase space, the cosmological constant   play the role of pressure  $P$ \cite{ref41,ref42}, and its thermal conjugate variable is the thermodynamic volume  $V$ of the black hole \cite{ref43,ref44}. The pressure and volume can be expressed as
\begin{align}
P=-\frac{\Lambda}{8\pi}=\frac{3}{8\pi l^2},\quad  V_h=\left(\frac{\partial M}{\partial P}\right)_{S,Q}=\frac{4\pi {r_h}^3}{3}.  \label{eq2.7}
\end{align}
The first law of thermodynamics thus should be written as \cite{ref50,ref51}
\begin{align}
dM=T_{h}dS_{h}+\Phi_{h}dQ+V_{h}dP.  \label{eq2.8}
\end{align}
In this case, the mass is defined as enthalpy. The relations among the enthalpy, internal energy and  pressure  are
\begin{align}
M=U_h+P V_h.  \label{eq2.9}
\end{align}
In the extended phase space, the change of mass will affect not only the event horizon and  the  electric charge, but also   the  $AdS$  radius. Therefore, we will investigate the change in the black hole by the charged particle absorption.

\section{Thermodynamic and Weak Cosmic Censorship Conjecture with contributions of pressure and volume}
\label{sec3}

\subsection{Energy-momentum relation of the absorbed particle}
 In order to obtain the relationship between the conserved quantities of particles in the electric  field $A_\mu$, we will employ  the following Hamilton-Jacobi equation to study the dynamical of the particles \cite{ref48}
\begin{align}
g^{\mu \nu }\left(p_{\mu }-\text{eA}_{\mu }\right)\left(p_{\nu }-\text{eA}_{\nu }\right)+\mu_b ^2=0,  \label{eq3.1}
\end{align}
where
\begin{align}
 p_\mu=\partial_\mu S.  \label{eq3.2}
\end{align}
In the above equation, $\mu_b$ is the mass of the particle, $p_\mu$ is the momentum, and $S$ is the Hamilton action of the particle. In the spherically symmetric spacetime,  the Hamilton action  of moving particle can be separated into
\begin{align}
S=-Et+R(r)+H(\theta )+L\phi,  \label{eq3.3}
\end{align}
here $E$ and $L$ are the energy and angular momentum. To solve the Hamilton-Jacobi equation,  we will use the inverse metric of the black hole as follows
\begin{align}
g^{\mu \nu } \partial _{\mu }\partial _{\nu }=-\frac{1}{f(r)}\left(\partial _t\right){}^2+f(r)\left(\partial _r\right){}^2+\frac{1}{r^2}\left(\partial _{\theta }\right){}^2+\frac{1}{r^2\sin ^2\theta }\left(\partial _{\phi }\right){}^2.  \label{eq3.4}
\end{align}
So the Hamilton-Jacobi equation changes into
\begin{align}
-\frac{1}{f(r)}\left(-E -\text{eA}_t\right){}^2+f(r)\left(\partial _rS(r)\right){}^2+\frac{1}{r^2}\left(\partial _{\theta }H(\theta )\right){}^2+\frac{1}{r^2\sin ^2\theta }L^2+\mu_b ^2=0.  \label{eq3.5}
\end{align}
Substituting equation (\ref{eq3.3}) into  equation (\ref{eq3.5}), we can get the radial  and angular equations
\begin{align}
-\frac{r^2}{f (r)} \left(-E -\text{eA}_t\right){}^2+r^2f (r) \left(\partial _rS(r)\right){}^2+r^2\mu_b ^2=-\mathcal{K},  \label{eq3.6}
\end{align}
\begin{align}
\left(\partial _{\theta }H(\theta )\right){}^2+\frac{1}{\sin ^2\theta }L^2=\mathcal{K}.  \label{eq3.7}
\end{align}
Correspondingly,  the radial momentum $p^r$ and angular momentum $p^{\theta }$ of the particle can be written as
\begin{align}
p^r=f(r)\sqrt{\frac{-\mu_b ^2r^2-\mathcal{K}}{r^2f(r)}+\frac{1}{f(r)^2}\left(-E -{eA}_t\right){}^2},  \label{eq3.8}
\end{align}
\begin{align}
p^{\theta }=\frac{1}{r^2}\sqrt{\mathcal{\mathcal{K}}-\frac{1}{\sin ^2\theta }L^2}.  \label{eq3.9}
\end{align}
 When the black hole absorbs a charged particle completely, the conserved quantity of the particle and the conserved quantity of the black hole are indistinguishable by an observer outside   the horizon. By removing the separate variable $\mathcal{K}$ in equation (\ref{eq3.8}), we can obtain the relationship between the energy  and momentum at any radial location. Near the event horizon, we can get
\begin{align}
E=\frac{Q}{r_h}e+p^r.  \label{eq3.10}
\end{align}
For the $p^r$ term, we will choose the positive sign thereafter as done in \cite{ref52} in order to assure a positive time direction.

\subsection{Thermodynamics in the extended phase space}

In the process of absorption, the energy and electric charge  of the particle equal  to the change of the internal energy and charge of the black hole, that is
\begin{align}
E=dU_h=d(M-P V_h),\quad e=dQ.  \label{eq3.11}
\end{align}
In this case, the energy relation of equation (\ref{eq3.10}) becomes as
\begin{align}
dU_h=\frac{Q}{r_h} dQ+p^r. \label{eq3.12}
\end{align}
In addition, in order to rewrite equation (\ref{eq3.12}) to the first law of thermodynamics, we have to find the variation of the entropy. From equation (\ref{eq2.5}), we have
\begin{align}
dS_h=2\pi r_h dr_h,  \label{eq3.13}
\end{align}
where $ dr_h$ is the variation of the event horizon of the black hole.  The  absorbed particles   also change the  function $f(r)$,  the shift of function $f(r)$  satisfies
\begin{align}
d{f_h}=\frac{\partial f_h}{\partial M}{dM}+\frac{\partial f_h}{\partial Q}{dQ}+\frac{\partial f_h}{\partial l}{dl}+\frac{\partial f_h}{\partial r_h}{dr}_h=0,\quad f_h=f\left(M,Q,l,r_h\right),  \label{eq3.14}
\end{align}
where
\begin{align}
&\frac{\partial f_h}{\partial M}=-\frac{2}{r_h}\text{   },  \quad  \frac{\partial f_h}{\partial Q}=\frac{2 Q}{r_h{}^2},\nonumber\\
&\frac{\partial f_h}{\partial l}=-\frac{2 r_h{}^2}{l^3},\quad  \frac{\partial f_h}{\partial r_h}=-\frac{2 Q^2}{r_h{}^3}+\frac{2 M}{r_h{}^2}+\frac{2 r_h}{l^2}-a r_h{}^{-2-3 w} (-1-3 w).  \label{eq3.15}
\end{align}
Combining equations (\ref{eq3.12}) and (\ref{eq3.11}), the energy relation in the equation (\ref{eq3.12}) can be rewritten as
\begin{align}
dM-d(PV_h)=\frac{Q}{r_h} dQ+p^r.  \label{eq3.16}
\end{align}
Combining equations (\ref{eq3.14}) and (\ref{eq3.16}), we find  that all the variables are eliminated except for $dr_h$ and $p^r$, so we can get
\begin{align}
{dr}_h=\frac{2 l^2 p^r r_h{}^{1+3 w}}{r_h{}^{3 w} \left(r_h{}^3-2 l^2 \left(M-r_h\right)\right)+a l^2 (3 w-1)}.  \label{eq3.17}
\end{align}
  Base on equation (\ref{eq3.17}), we can get the variation of entropy and volume
\begin{align}
{dS}_h=\frac{4 l^2 \pi p^r r_h{}^{2+3 w}}{r_h{}^{3 w} \left(r_h{}^3-2 l^2 \left(M-r_h\right)\right)+a l^2 (3 w-1)},  \label{eq3.18}
\end{align}
\begin{align}
{dV}_h=\frac{8 l^2 \pi  p^r r_h{}^{3+3 w}}{r_h{}^{3 w} \left(r_h{}^3-2 l^2 \left(M-r_h\right)\right)+a l^2 (3 w-1)}.   \label{eq3.19}
\end{align}
Incorporating equations (\ref{eq2.4}), (\ref{eq3.18}), (\ref{eq2.7}) and (\ref{eq3.19}), we  get lastly
\begin{align}
 T_h {dS}_h-{PdV}_h=p^r.  \label{eq3.20}
\end{align}
Moreover, incorporating equations (\ref{eq2.4}), (\ref{eq2.5}), (\ref{eq2.6}), (\ref{eq2.7}), (\ref{eq3.18}) and (\ref{eq3.19}), the energy relation of equation (\ref{eq3.12}) can be expressed as
\begin{align}
dU_h=\Phi_h dQ + T_h dS_h-P dV_h.  \label{eq3.21}
\end{align}
Because the mass of the black hole has been defined as enthalpy, the relation between the  internal energy and enthalpy can be rewritten  in the extended phase space as
\begin{align}
dM=T_h dS_h+\Phi_h dQ+V_h dP.  \label{eq3.22}
\end{align}
Here, we prove that the first law of thermodynamics is still satisfied  for the black hole surrounded by quintessence dark energy under the charged particle absorption.

As the absorption is an irreversible process, the entropy of the final state  should be greater than the initial state of the black hole. That is, under the charged particle absorption, the variation of the entropy is $dS > 0$. Next, we will test the validity of the second law of thermodynamics with equation (\ref{eq3.18}).

We first study the case of the extremal black holes, for which  the temperature   is zero. On the basis of this fact and equation (\ref{eq3.18}), we can get
\begin{align}
dS_h=-\frac{4\pi p^r l^2}{3r_h}.  \label{eq3.23}
\end{align}
There is a minus sign in   equation (\ref{eq3.23}). Therefore, the entropy is decreased  for the extremal  black hole. That is, the second law of thermodynamics is violated for the extremal black hole in extended phase space. In addition, it is worth noting that the parameters $a$ and  $\omega$ are not present in the above equation. In other words, the violations about the second law does not depend on the parameters $a$ and  $\omega$.

Now, we focus on the near-extremal black hole.  We will check whether the second law   is valid in the extended phase space by studying the variation of entropy numerically. We set $M = 0.5$ and $l = p^r = 1$. For the case $\omega=-2/3$ and $a=1/3$, we find the extremal charge is $Q_e =0.48725900857$. In the case that the charge is less than the extremal charge, we take different charge values to produce the variation of entropy. In Table 1,  we give the  numeric results of $r_h$ and $dS$ for different charges. As can be seen from Table 1, when the charge $Q$ of the black hole decreases, the event horizon of the black hole increases, and the variation  of entropy increase too. Interestingly, there are two regions where the  entropy increase are $dS>0$ and $dS<0$. It means that there exist a phase transition point  which divides the variation of entropy into positive value and negative value.
\begin{center}
{\footnotesize{\bf Table 1.} The relation between $dS_h$, $Q$ and $r_h$.\\
\vspace{2mm}
\begin{tabular}{ccc}
\hline
$Q $               &$r_h $             & $dS_h $         \\
\hline
0.48725900857     & 0.432041           & $-9.69538$     \\
0.487259          & 0.432111           & $-9.70328$     \\
0.48              & 0.494855           & $-22.9663 $    \\
0.46              & 0.567768           & $-1006.44$     \\
0.445             & 0.5757             & $386.566$      \\
0.4               & 0.628743           & $44.8667 $     \\
0.3               & 0.695449           & $24.9494 $     \\
0.2               & 0.731872           & $21.0262 $     \\
0.1               & 0.751055           & $19.5985$      \\
\hline
\end{tabular}}
\end{center}

 We also can obtain the relation between $dS$ and $r_h$, which is shown in Figure 1. From Figure 1, we can clearly see that there really exist a phase transition point making $dS_h$ positive and negative. When the electric charge is close to the extremal charge, the  variation of entropy   is negative. And when the electric charge is far away from the extremal charge, the entropy  increases.  Therefore,  the second law of thermodynamics is violated for the near-extremal black holes and valid for the far-extremal black holes in the extended phase space.

\begin{figure}[htb]
\centering
\includegraphics[scale=0.8]{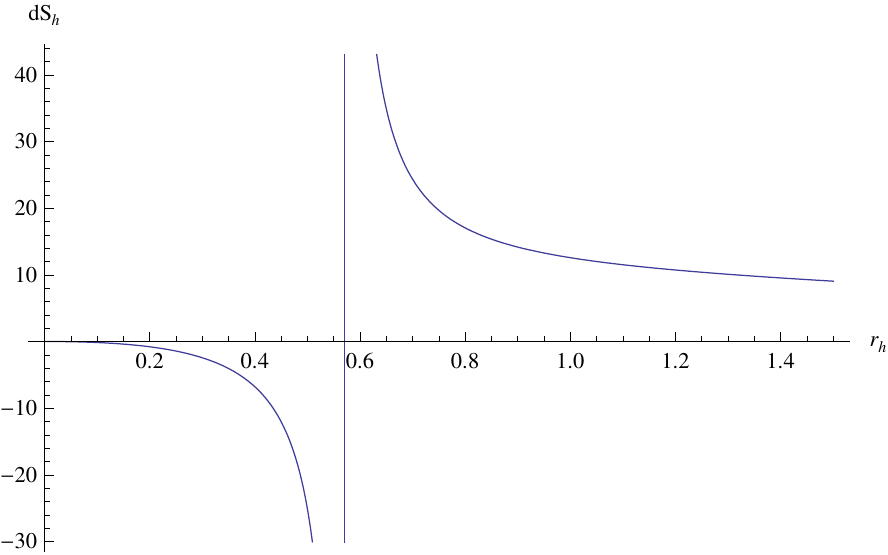}
\caption{The relation between $dS_h$ and $r_h$ which parameter values are $M = 0.5, l = p^r = 1$ and $\omega=-2/3, a=1/3$.}
\label{fig:1}
\end{figure}
In   Figure 2 and Table 2, we set  $ \omega= -1/2$ and $a = 1/2$. We want to explore whether the values of state parameter of dark energy affect the laws of thermodynamics. For $ \omega= -1/2$ and $a = 1/2$, the extremal charge is $Q_e = 0.525694072$. Form Figure 2 and Table 2, we also find that the second law of thermodynamics fails for the  near-extremal black holes when a particle is absorbed by the black hole.   In addition, by comparing Figure 1 and Figure 2, we find that the magnitudes of the violations are related to values of the parameters $ \omega$ and $a$,  but  parameters $ \omega$ and $a$ do not determine whether the second law of thermodynamics will eventually be violated. 
\begin{center}
{\footnotesize{\bf Table 2.} The relation between $dS_h$, $Q$ and $r_h$.\\
\vspace{2mm}
\begin{tabular}{ccc}
\hline
$Q $               &$r_h $             & $dS_h $         \\
\hline
0.525694072      & 0.483844         & $-8.65919 $    \\
0.525            & 0.504022         & $-10,2582$    \\
0.5              & 0.599642         & $-27.2971 $    \\
0.43             & 0.69007          & $-766.564$      \\
0.425            & 0.694382         & $20923.7$      \\
0.4              & 0.713732         & $171.512 $     \\
0.3              & 0.767156         & $50.6936 $     \\
0.2              & 0.798062         & $37.5343 $     \\
0.1              & 0.814681         & $33.2405$     \\
\hline
\end{tabular}}
\end{center}

\begin{figure}[htb]
\centering
\includegraphics[scale=0.8]{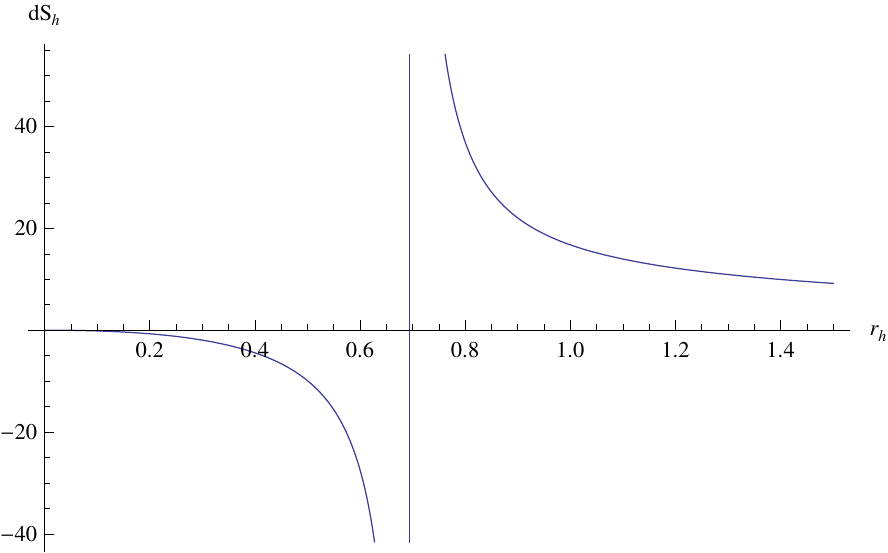}
\caption{The relation between $dS_h$ and $r_h$ which parameter values are $M = 0.5, l = p^r = 1$ and $ \omega= -1/2$, $a = 1/2$.}
\label{fig:1}
\end{figure}

\subsection{Weak Cosmic Censorship Conjecture in the extended phase space}

 In the extended phase space with consideration of the thermodynamic volume,  although the particle absorption is an irreversible process, the second law of thermodynamics for the extremal and near-extremal black holes are violated.  As well as known, the definition of entropy is $S_h=\frac{A_h}{4}$,  the event horizon of black holes is closely relevant to entropy. Duo to  this  violation of the second law of the thermodynamics is found in the case considering thermodynamic volume. It implies that, the violation of the second law of thermodynamics can be related to the weak cosmic censorship conjecture which is related to the stability of the event horizon. Therefore, it is necessary to check the validity of the weak cosmic censorship conjecture in these cases. 

  If the event horizon cannot wrap the singularity   after the charged particle are absorbed, the weak cosmic censorship conjecture will be invalid.  So the event horizon should exist to assure the validity of the weak cosmic censorship conjecture.  We will check whether there is an event horizon after the charged particle is absorbed by the black hole.  We will pay attention to how $f(r)$ changes. The function $f(r)$ has a minimum point at $r_\text{min}$.  There are three possibilities. For the case $f(r)_\text{min }<0$,  there are two roots of $f(r)$. Then we have an usual black hole with ${r_+}$ and ${r_-}$, as the inner  horizon and outer horizon. For the case $f(r)_\text{min }=0$, the two roots coincide, and the black hole becomes into an extremal black hole. For the case $f(r)_\text{min }>0$, the function $f(r)$ has no real root,  there is not an event horizon. At $r_\text{min}$,  we have
\begin{align}
f(M,Q,l,r)|_{r=r_\text{min}}\equiv f_\text{min}=\delta\leq 0,\quad \partial_{r} f(M,Q,l,r)|_{r=r_\text{min}}\equiv f'_\text{min}=0,  \label{eq3.24}
\end{align}
and
\begin{align}
 (\partial_{r})^2 f(M,Q,l,r)|_{r=r_\text{min}}>0.  \label{eq3.25}
\end{align}
 For the extremal black hole, $\delta=0$, $r_h$ and $r_\text{min }$ are coincident. For the near-extremal black hole, $\delta$ is a small quantity. When the black hole absorbs charged particle, there are infinitesimal change  in the mass, charge and AdS radius, which are $(M+dM,Q+dQ,l+dl)$. Because of  these changes, there are also  movements for the minimum value and event horizon of the black hole, namely $r_\text{min}\rightarrow r_\text{min}+dr_\text{min}$ and $r_h\rightarrow r_h+dr_h$. At the new minimum point, the function $f(r)$  satisfies
\begin{align}
\partial_{r} f(M+dM,Q+dQ,l+dl,r)|_{r=r_\text{min}+dr_\text{min}}=f'_\text{min}+df'_\text{min}=0.  \label{eq3.26}
\end{align}
Using the known condition $f'_\text{min}=0$, we can obtain $df'_\text{min}=0$. Expanding it further, we get
\begin{align}
df'_\text{min}=\frac{\partial f'_\text{min}}{\partial M}dM+\frac{\partial f'_\text{min}}{\partial Q}dQ+\frac{\partial f'_\text{min}}{\partial l}dl+\frac{\partial f'_\text{min}}{\partial r_\text{min}}dr_\text{min}=0.  \label{eq3.27}
\end{align}
 At   $r_\text{min}+dr_\text{min}$,  the function $f(r)$ takes the form as
\begin{align}
f(M+dM,Q+dQ,l+dl,r)&|_{r=r_\text{min}+dr_\text{min}}=f_\text{min}+df_\text{min}\nonumber\\
&=\delta+\left(\frac{\partial f_\text{min}}{\partial M}dM+\frac{\partial f_\text{min}}{\partial Q}dQ+\frac{\partial f_\text{min}}{\partial l}dl\right).  \label{eq3.28}
\end{align}
For extremal black hole, we know  $f'_\text{min}=0$ and $ f_\text{min }=\delta=0 $.  Substituting equation (\ref{eq3.16}) into equation (\ref{eq3.28}),  we get
\begin{align}
df_{\min }= -\frac{2 p^r}{r_{\min }}-\frac{3 r_{\min } dr_{\min } }{l^2}. \label{eq3.29}
\end{align}
Form equations (\ref{eq3.20}) and (\ref{eq3.29}), we find
\begin{align}
df_{\min }=0.  \label{eq3.30}
\end{align}

For the near-extremal black hole, equation (\ref{eq3.16}) can not be applicable  no longer. But we can expand it   near the minimum  point for there is a relation $r_h=r_\text{min }+\epsilon $.    To the first order,  we find
\begin{align}
&dM=\frac{r_{\min }^{-2-3 \omega } {dr}_{\min }\left({al}^3 r_{\min }+3 a l^3 \omega  r_{\min }-2 l^3 Q^2 r_{\min }^{3 \omega }+2 l^3 M r_{\min }^{1+3 \omega }+2 l r_{\min }^{4+3 \omega }\right)}{2 l^3} \nonumber\\
&~~~~~+\frac{r_{\min }^{-2-3 \omega }\left(2 l^3 Q r_{\min }^{1+3 \omega }{dQ}-2r_{\min }^{5+3 \omega }{dl}\right)}{2 l^3} \nonumber\\
&~~~~~-\frac{r_{\min }^{-3-3 \omega }{dr}_{\min } \left(a l^3 r_{\min }+6 a l^3 \omega  r_{\min }+9 a l^3 \omega ^2 r_{\min }\right) \epsilon }{2 l^3} \nonumber\\
&~~~~~-\frac{r_{\min }^{-3-3 \omega }{dr}_{\min } \left(-4 l^3 Q^2 r_{\min }^{3 \omega }+2 l^3 M r_{\min }^{1+3 \omega }-4 l r_{\min }^{4+3 \omega }\right) \epsilon }{2 l^3} \nonumber\\
&~~~~~-\frac{r_{\min }^{-3-3 \omega }\left(2 l^3 Q r_{\min }^{1+3 \omega }{dQ}+6 r_{\min }^{5+3 \omega }{dl}\right) \epsilon }{2 l^3}+\mathcal O(\epsilon )^2.  \label{eq3.31}
\end{align}
Substituting (\ref{eq3.31}) into (\ref{eq3.28}), we have
\begin{align}
&{df}_{\min }=\frac{2 Q {dQ}}{r_{\min }^2}-\frac{2 r_{\min }^2{dl}}{l^3}-\frac{r_{\min }^{-3-3\omega } \left(2 l^3 Q r_{\min }^{1+3 \omega }{dQ}-2 r_{\min }^{5+3\omega }{dl}\right)}{l^3} \nonumber\\
&~~~~~-\frac{r_{\min }^{-3-3 \omega }{dr}_{\min }\left({al}^3r_{\min }+3a l^3\omega r_{\min }-2 l^3 Q^2 r_{\min }^{3 \omega }+2 l^3 M r_{\min }^{1+3 \omega }+2 l r_{\min }^{4+3 \omega }\right)}{l^3} \nonumber\\
&~~~~~+\frac{r_{\min }^{-4-3 \omega } {dr}_{\min }\left(a l^3 r_{\min }+6 a l^3 \omega r_{\min }+9 a l^3 \omega ^2 r_{\min }\right) \epsilon }{l^3} \nonumber\\
&~~~~~+\frac{r_{\min }^{-4-3 \omega } {dr}_{\min }\left(-4 l^3 Q^2 r_{\min }^{3 \omega }+2 l^3 M r_{\min }^{1+3 \omega }-4 l r_{\min }^{4+3 \omega }\right) \epsilon }{l^3} \nonumber\\
&~~~~~+\frac{r_{\min }^{-4-3 \omega } \left(2 l^3 Q r_{\min }^{1+3 \omega }{dQ} +6 r_{\min }^{5+3 \omega } {dl}\right) \epsilon }{l^3}+\mathcal O(\epsilon )^2.  \label{eq3.32}
\end{align}
For the equation $f(r_h)=0$, we also can expand and solve it, then we find
\begin{align}
l=\frac{\sqrt{3} r_{\min }^{\frac{1}{2} (4+3 \omega )}}{\sqrt{-3 a \omega  r_{\min }+Q^2 r_{\min }^{3 \omega }-r_{\min }^{2+3 \omega }}},  \label{eq3.33}
\end{align}
Differentiating both sides of this equation, we can get further
\begin{align}
&{dl}=-\frac{\sqrt{3}\text{  }Q r_{\min }^{3 \omega +\frac{1}{2} (4+3 \omega )}{dQ}}{\left(-3 a \omega  r_{\min }+Q^2 r_{\min }^{3 \omega }-r_{\min }^{2+3 \omega }\right){}^{3/2}} \nonumber\\
&~~~~~+\frac{\sqrt{3} (4+3 \omega ) r_{\min }^{-1+\frac{1}{2} (4+3 \omega )}{dr}_{\min }}{2 \sqrt{-3 a \omega  r_{\min }+Q^2 r_{\min }^{3 \omega }-r_{\min }^{2+3 \omega }}} \nonumber\\
&~~~~~-\frac{\sqrt{3} r_{\min }^{\frac{1}{2} (4+3 \omega )} \left(-3 a \omega +3 Q^2 \omega  r_{\min }^{-1+3 \omega }-(2+3 \omega ) r_{\min }^{1+3 \omega }\right){dr}_{\min }}{2 \left(-3 a \omega  r_{\min }+Q^2 r_{\min }^{3 \omega }-r_{\min }^{2+3 \omega }\right){}^{3/2}}.  \label{eq3.34}
\end{align}
Substituting equations (\ref{eq3.34}) and (\ref{eq3.33}) into equation (\ref{eq3.32}), we find
\begin{align}
df_{\min }=\mathcal O(\epsilon^2 ),  \label{eq3.35}
\end{align}
then, we can get
\begin{align}
f_\text{min}+df_\text{min}=\delta +\mathcal O(\epsilon^2 ). \label{eq3.36}
\end{align}
 In Ref. \cite{Gwak:2017kkt},  $\delta $ is a very small negative value and $\epsilon\ll1$. When we chose  $\delta =\epsilon=0$, the equation (\ref{eq3.36}) return to the form of the extremal black hole. This result also proves the correctness of the equation (\ref{eq3.30}) from the side.  Hence, they think that the extremal and near-extremal black holes stay at their minimum. For the near-extremal black hole, they ignore the $O(\epsilon^2 )$ due to it is high order small value, so that they think that the weak cosmic censorship is also valid for the case of the near-extremal black hole. However, because  we don't know the magnitude of $|\delta |$ and $ O(\epsilon^2 )$, we can not easily ignore  $ O(\epsilon^2 )$. Therefore, it would be important to find out whether there is a relationship $|\delta |>\mathcal O(\epsilon^2 )$.
For the near-extremal black hole, the  minimum point is located slightly to the left of the outer horizon. So the value of the minimum is obtained at $r_\text{min}+\epsilon$ as
\begin{align}
&f{\left(r_\text{min}+\epsilon \right)}=\left(1-\frac{Q^2}{r_\text{min}^2}+\frac{3 r_\text{min}^2}{l^2}+3 a \omega  r_\text{min}^{-1-3 \omega }\right)  \nonumber\\
&~~~~~~~~~~~~~~+\left(\frac{3}{l^2}+\frac{1}{2} r_\text{min}^{-4-3 \omega } \left(-3 a \omega  (1+3 \omega ) r_\text{min}+2 Q^2 r_\text{min}^{3 \omega }\right)\right) \epsilon ^2+\mathcal O(\epsilon^3), \label{eq3.37}
\end{align}
where we skip to write the constant term. Namely
\begin{align}
\delta =-\left(\frac{3}{l^2}+\frac{1}{2} r_\text{min}^{-4-3 \omega } \left(-3 a \omega  (1+3 \omega ) r_\text{min}+2 Q^2 r_\text{min}^{3 \omega }\right)\right) \epsilon ^2-\mathcal O(\epsilon^3). \label{eq3.38}
\end{align}
And we can also get
\begin{align}
df\text{min}=\left(\frac{\left(8 Q^2 -2{  }r_{\min }^2\right){dr}_{\min }-4 Q r_{\min }{dQ}}{r_{\min }^5}-\frac{27}{2} a \omega  (1+\omega )^2{  }r_{\min }^{-4-3 \omega }{dr}_{\min }\right) \epsilon ^2+\mathcal O(\epsilon^3). \label{eq3.39}
\end{align}
the equation (\ref{eq3.36}) can be represented as
\begin{align}
&f{\left(r_\text{min}+dr_\text{min} \right)}=\left(\frac{\left(8 Q^2 -2\text{  }r_{\min }^2\right){dr}_{\min }-4 Q r_{\min }{dQ}}{r_{\min }^5}-\frac{27}{2} a \omega  (1+\omega )^2{  }r_{\min }^{-4-3 \omega }{dr}_{\min }\right)\epsilon ^2  \nonumber\\
&~~~~~~~~~~~~~~~~~-\left(\frac{3}{l^2}+\frac{1}{2} r_\text{min}^{-4-3 \omega } \left(-3 a \omega  (1+3 \omega ) r_\text{min}+2 Q^2 r_\text{min}^{3 \omega }\right)\right) \epsilon ^2. \label{eq3.40}
\end{align}
Here, we can have $\mathcal F=\frac{f{\left(r_{\min }+\text{dr}_{\min }\right)}}{\epsilon ^2}$ and $e=dQ$. Then, the above equation becomes
\begin{align}
&\mathcal F=\frac{\left(8 Q^2 -2{  }r_{\min }^2\right){dr}_{\min }-4 Q r_{\min }e}{r_{\min }^5} -\frac{27}{2} a \omega  (1+\omega )^2{  }r_{\min }^{-4-3 \omega }{dr}_{\min }\nonumber\\
&~~~~~~~-\frac{1}{2} r_{\min }^{-4-3 \omega } \left(-3 a \omega  (1+3 \omega ) r_{\min }+2 Q^2 r_{\min }^{3 \omega }\right)-\frac{3}{l^2}, \label{eq3.41}
\end{align}
 obviously, from above equation we find that  the values of  $\mathcal F$ is link with  the values of the parameters $e, a, l, r_{\min}, \omega, Q, dr_{\min}$. Mostly, we find that $e$ has a more direct affect on  $\mathcal F$. To gain an intuitive understanding, we  plot Figure 3  in follow 
\begin{figure}[htb]
 {\resizebox{0.3\textwidth}{!}{\includegraphics {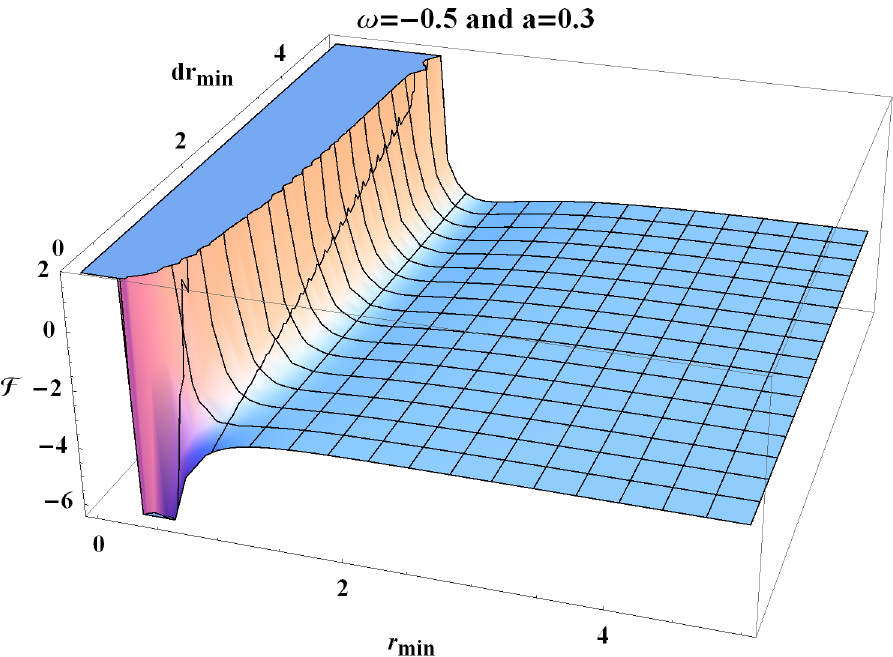}}}
  {\resizebox{0.3\textwidth}{!}{\includegraphics{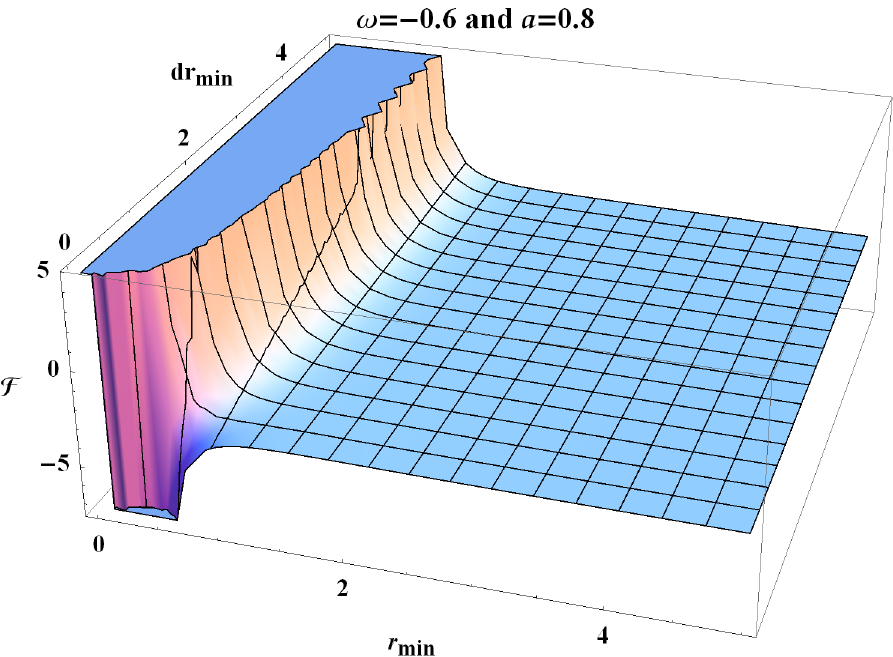}}}
  {\resizebox{0.3\textwidth}{!}{\includegraphics{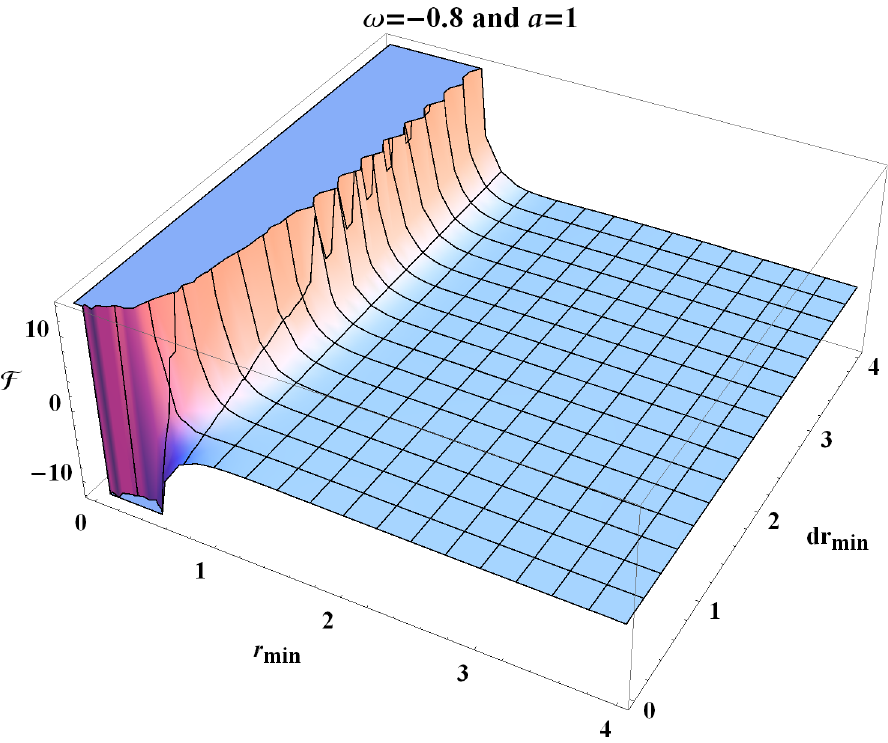}}}
 \caption{The value of  $\mathcal F$ for $l=1, Q=0.5, e=0.2$.}
\label{fig:3}
\end{figure}
\begin{figure}[htb]
 {\resizebox{0.3\textwidth}{!}{\includegraphics{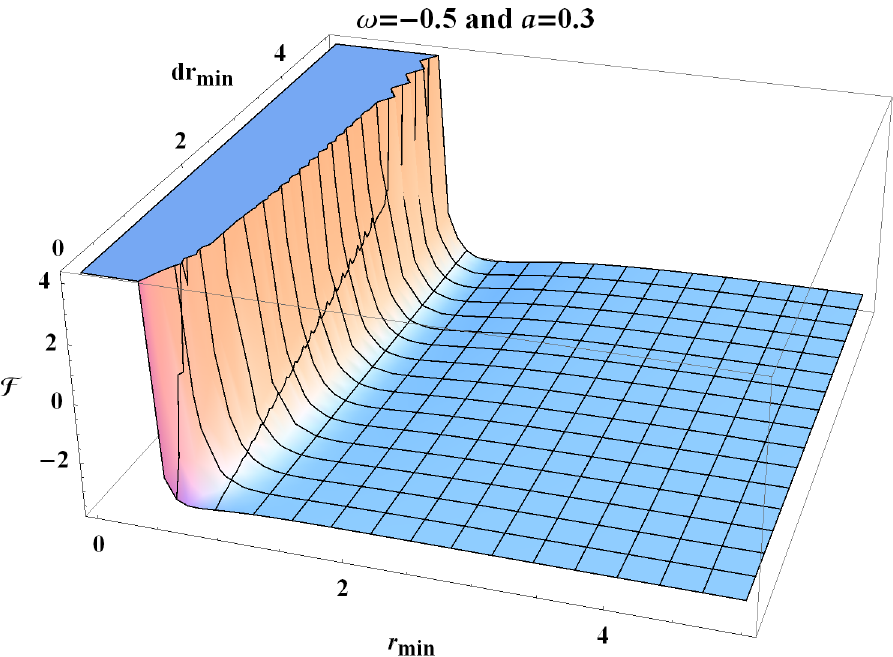}}}
  {\resizebox{0.3\textwidth}{!}{\includegraphics{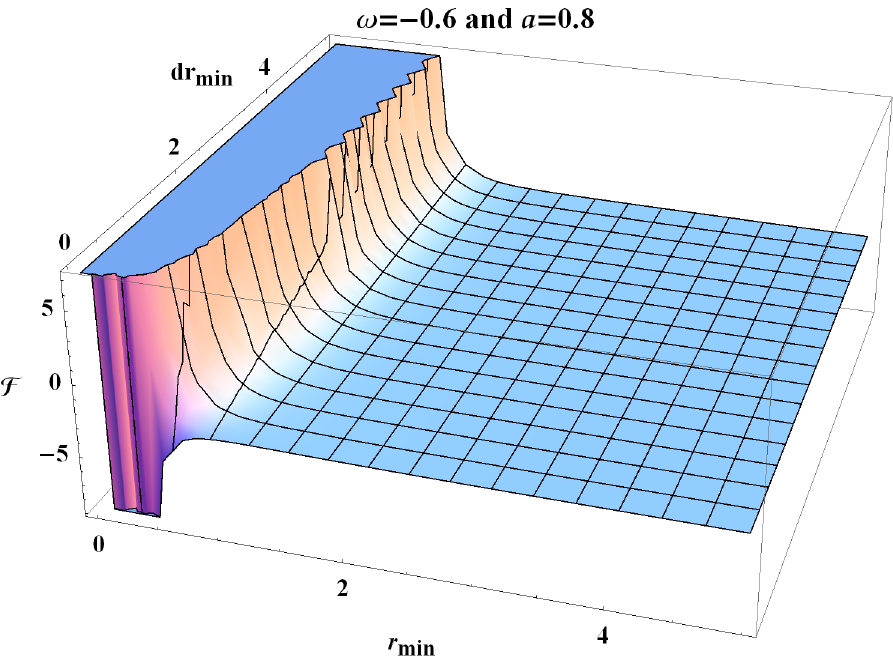}}}
  {\resizebox{0.3\textwidth}{!}{\includegraphics{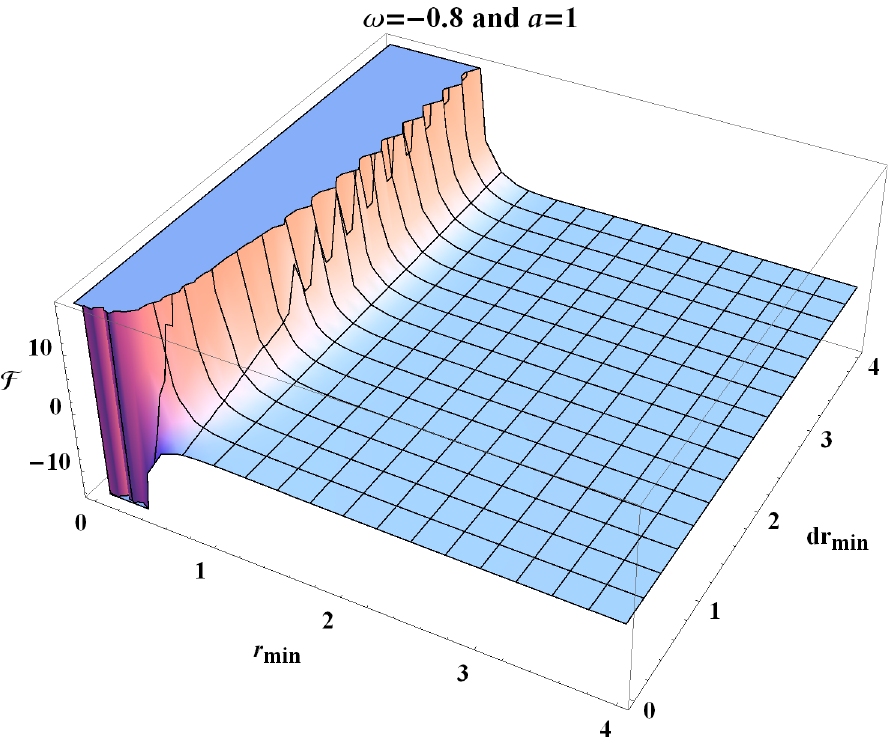}}}
 \caption{The value of  $\mathcal F$ for $l=1, Q=0.5, e=0.00005$.}
\label{fig:3}
\end{figure}

When we take different values of these parameters, the region where $\mathcal F$ is greater than zero always appears in the graph.  $\mathcal F>0$, that is,  $f{\left(r_\text{min}+dr_\text{min} \right)}>0$.  In other words,  the positive region shows that there exists a rang of the particle charge which allow us to overcharge black holes into  naked singularities.  Similarly, for different values of the parameters $a, l, r_{\min}, \omega, Q, dr_{\min}$, the configurations of $\mathcal F$ are different, that is, the violation about the weak cosmic censorship depend on the parameters and the magnitudes of the violation is related to those of the parameters. Especially,  when we take $e<<1$, and show in the Figure 4, we still find the same violation. However, the degree of violation is different as the change of the charge $e$

In the extended phase space, we find  that the change of the   values  of $f(r_\text{min})$  vanishes always after the  charged particle is absorbed for the extremal black hole. That is,  for the extremal black holes, the function $f(r)$ always has a root. Therefore, the black hole has an event horizon covering its singularity. The weak cosmic censorship conjecture thus is valid for the configuration of the black hole does  not change. However, we derive that the near-extremal black holes can be overcharged by absorbing charged particle. Hence, the cosmic censorship conjecture would be violated for the near-extremal black holes in the extended phase space.

 When $a$ is zero, the black hole becomes RN-AdS black holes. Hence, we set $a=0$, and the result is shown in Figure 5. 

\begin{figure}[htb]
 {\resizebox{0.3\textwidth}{!}{\includegraphics{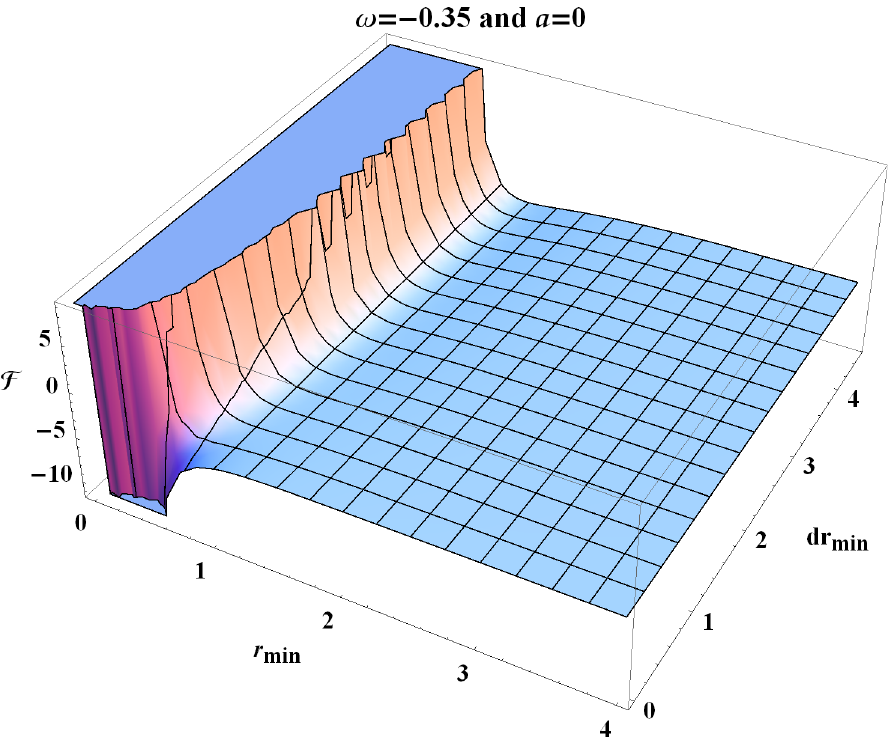}}}
  {\resizebox{0.3\textwidth}{!}{\includegraphics{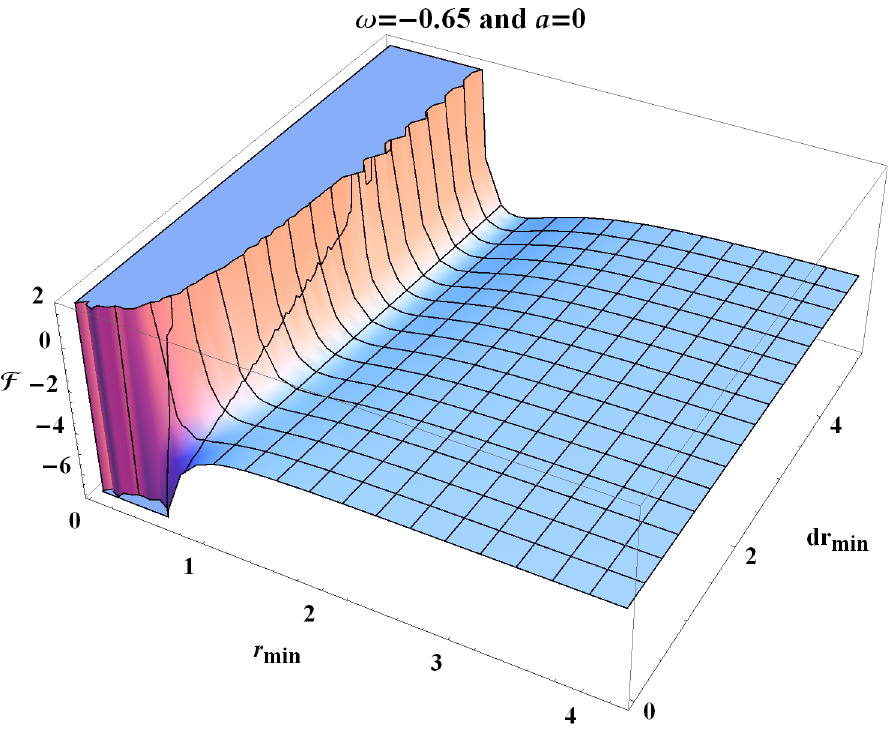}}}
  {\resizebox{0.3\textwidth}{!}{\includegraphics{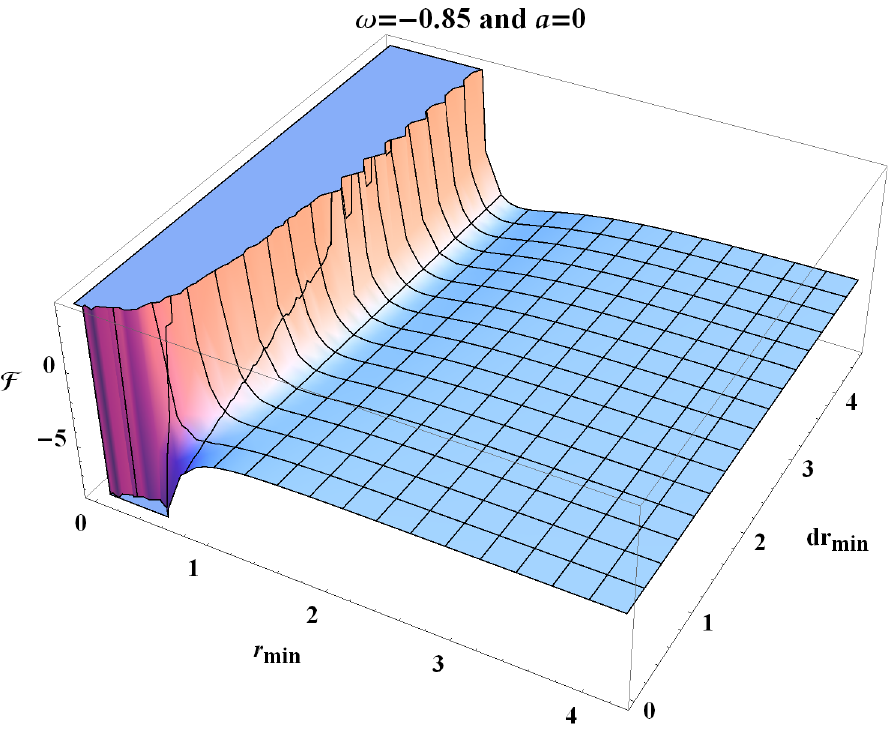}}}
 \caption{The value of  $\mathcal F$ for $l=1, Q=0.5, e=0.2$.}
\label{fig:3}
\end{figure}

 {Interestingly, the results show that not only   $\mathcal F<0$ but also   $\mathcal F>0$, which means the weak cosmic censorship may be invalid for the  RN-AdS black holes, and the results are some different  with the previous works \cite{Gwak:2017kkt} about the weak cosmic censorship. The reasons is that they did not consider the high order corrections to the energy.  

\section{Thermodynamics and the weak cosmic censorship conjecture without  contributions of pressure and volume}
\label{sec4}
In this section, we mainly discuss the laws of thermodynamics    and examine the weak cosmic censorship conjecture of the black hole in the normal phase space. We want to  explore whether the phase space affects thermodynamics and weak cosmic censorship conjecture.

\subsection{Thermodynamic in the normal phase space}

In the normal phase space,  the mass $M$ of the black hole is defined as energy. Meanwhile, we assume that there is no energy loss in the process of particle absorption.  As  the charged particle is absorbed by the black hole, the change of the internal energy and charge of the black hole  satisfy
 \begin{align}
E=dM, \quad e=dQ.  \label{eq4.1}
\end{align}
Then, equation (\ref{eq3.10}) can be rewritten as
\begin{align}
dM=\frac{Q}{r_h} dQ+p^r.  \label{eq4.2}
\end{align}
Due to the absorption of a  charged particle, the event horizon and  function $f(r_h)$ will change,  and there is always a relation
\begin{align}
df_h=\frac{\partial f_h}{\partial M}{dM}+\frac{\partial f_h}{\partial Q}{dQ}+\frac{\partial f_h}{\partial r_h}{dr}_h=0.  \label{eq4.3}
\end{align}
In order to eliminate the $dM$ term, we can combine equations (\ref{eq4.2}) and (\ref{eq4.3}). We find that the $dQ$ term is also eliminated in this process and we get lastly
\begin{align}
{dr}_h=\frac{2 l^2 p^r r_h{}^2}{2l^2 r_h{}^2-2Ml^2 r_h+4r_h{}^4+a l^2r_h{}^{-(3\omega -1)} (3\omega -1)}.  \label{eq4.4}
\end{align}
After substituting equation (\ref{eq4.4}) into equation (\ref{eq3.13}), we have
\begin{align}
{dS}_h=\frac{4 \pi  l^2 p^r r_h^3}{2l^2 r_h{}^2-2Ml^2 r_h+4r_h{}^4+a l^2 r_h{}^{-(3\omega -1)} (3\omega -1)}.  \label{eq4.5}
\end{align}
Form equations (\ref{eq2.4}) and (\ref{eq4.5}), we get
\begin{align}
T_h {dS}_h=p^r.  \label{eq4.6}
\end{align}
Combining equations (\ref{eq2.4}), (\ref{eq2.6}) and (\ref{eq4.5}), we get
\begin{align}
{dM}=T_h {dS}_h+\Phi _h{dQ}.  \label{eq4.7}
\end{align}
 In the normal phase space, the first law of thermodynamics of the black hole surrounded by quintessence dark energy is valid when a particle is absorbed  by the black hole.

With  equations (\ref{eq4.5}), we also can investigate  the  second law of thermodynamics. For the extremal black holes, we find the variation of the entropy is divergent, which is meaningless.  So we are interested in the near-extremal black hole thereafter.   We also set $M = 0.5$ and $l = p^r = 1$. For case of $\omega=-1/2$ and $a=1/2$, we get the extremal charge $Q_e =0.525694072 $. In the case that the charge is  less than the extreme charge,  we obtain values of $r_h$ and $dS_h$ for different charge  in Table 3. From this table, it can be clearly seen that when the charge is smaller than the extremal charge, the variation of the entropy is positive always. In Figure 6, we present the relation between $dS_h$ and $r_h$,  we can see that the entropy increases too. Therefore, the second law of thermodynamics is valid  in the normal phase space.
\begin{center}
{\footnotesize{\bf Table 3.} The relation between $\text{dS}_h$, $Q$ and $r_h$.\\
\vspace{2mm}
\begin{tabular}{ccc}
\hline
$Q $               &$r_h $             & $dS_h $         \\
\hline
0.525694072      & 0.483844         & $39998.1 $    \\
0.525            & 0.504022         & $43.7772$    \\
0.5              & 0.599642         & $9.3879 $    \\
0.4              & 0.713732         & $5.67468$     \\
0.3              & 0.767156         & $4.92923 $     \\
0.2              & 0.798062         & $4.60478 $     \\
0.1              & 0.814681         & $4.45287$     \\
\hline
\end{tabular}}
\end{center}

\begin{figure}[htb]
\centering
\includegraphics[scale=0.8]{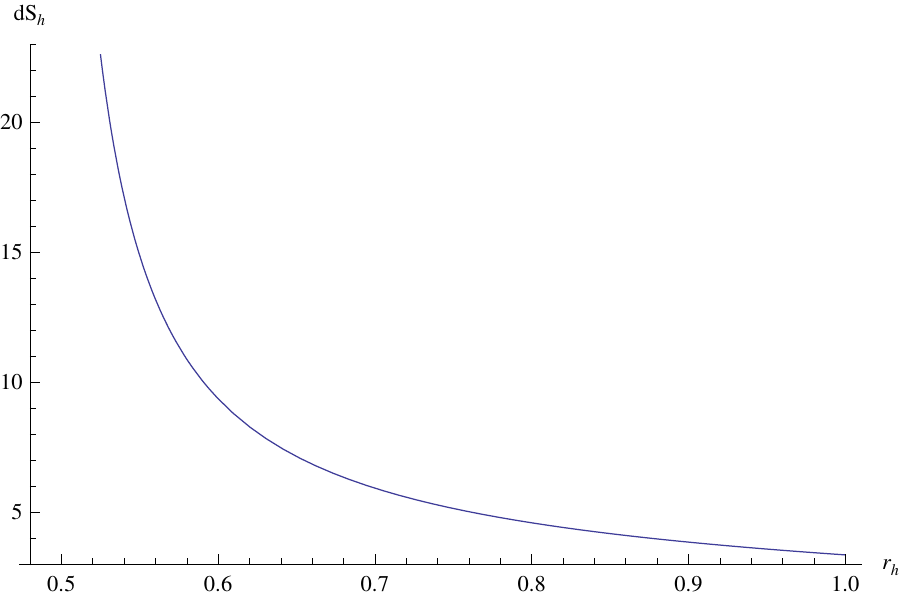}
\caption{The relation between $dS_h$ and $r_h$ which parameter values are $a=1,  M = 0.5, l = p^r = 1$.}
\label{fig:6}
\end{figure}

For the case  $\omega=-2/3, a=1/3$, we find the extremal charge is $Q_e =0.48725900875 $.  the values of $r_h$ and $dS_h$ for different charge are  given in Figure 7 and Table 4 From them, we also find that the entropy increase, implying that the second law  is valid.
\begin{center}
{\footnotesize{\bf Table 4.} The relation between $dS_h$, $Q$ and $r_h$.\\
\vspace{2mm}
\begin{tabular}{ccc}
\hline
$Q $               &$r_h $             & $dS_h $         \\
\hline
0.48725900875    & 0.432041         & $3.47995\times 10^6 $    \\
0.487259         & 0.434211         & $322.509$    \\
0.48725          & 0.434319        & $307.346 $    \\
0.4              & 0.628743         & $5.80081$     \\
0.3              & 0.695450         & $4.85183 $     \\
0.2              & 0.731837         & $4.49911 $     \\
0.1              & 0.751055         & $4.34168$     \\
\hline
\end{tabular}}
\end{center}
\begin{figure}[htb]
\centering
\includegraphics[scale=0.8]{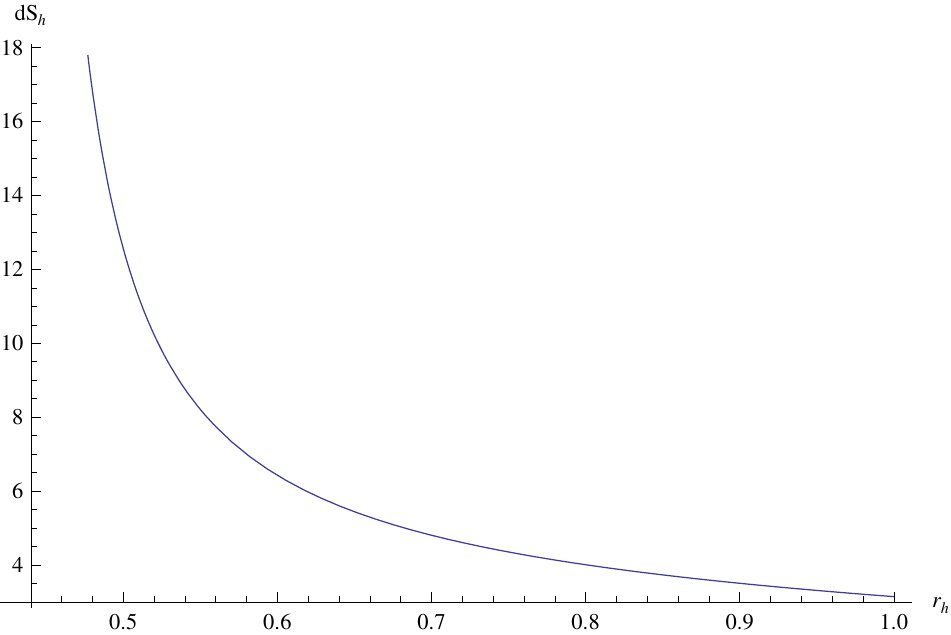}
\caption{The relation between $dS_h$ and $r_h$ which parameter values are $a=1/3,  M = 0.5, l = p^r = 1$.}
\label{fig:7}
\end{figure}
Thus far, both the first and second laws of thermodynamics hold  in the normal phase space for the black hole surrounded by quintessence dark energy under charged particle absorption.

\subsection{Weak Cosmic Censorship Conjecture in the normal phase space}

In the normal phase, the examination of the validity of the weak cosmic censorship conjecture  should also return to the value of the function $f(r_\text{min })$. Similarly, we will study how $f(r_\text{min })$ changes as charged particle are absorbed. At $r_\text{min }+{dr}_\text{min }$, there is also a relation $\partial _r f(r_\text{min }+{dr}_\text{min })=0$, implying
\begin{align}
{df}'_\text{min }=\frac{\partial f'_\text{min }}{\partial M} dM+\frac{\partial f'_\text{min }}{\partial Q} dQ+\frac{\partial f'_\text{min }}{\partial r_\text{min }} dr_\text{min }=0.  \label{eq4.8}
\end{align}
In addition, at the new minimum point, $ f\left(r_\text{min }+dr_\text{min }\right)$ can be expressed as
\begin{align}
f\left(r_\text{min }+dr_\text{min }\right)=f_\text{min }+df_\text{min }, \label{eq4.9}
\end{align}
where
\begin{align}
df_\text{min }=\frac{\partial f_\text{min }}{\partial M} dM+\frac{\partial f_\text{min }}{\partial Q} dQ. \label{eq4.10}
\end{align}
For the extremal black holes,  equation (\ref{eq4.2}) can be applied. In this case, we have $f_\text{min }=\delta =0$, inserting equation (\ref{eq4.2}) into equation (\ref{eq4.10}), we can finally get
\begin{align}
df_\text{min }=-\frac{2 p^r}{r_\text{min }}. \label{eq4.11}
\end{align}
When it is an extreme black hole, $r_h$ and $r_\text{min}$ are tightly coincident. In addition,  we have $T_h = 0$.  Incorporating equation (\ref{eq4.11}) and equation (\ref{eq4.6}), the minimum value of $f\left(r_\text{min }+dr_\text{min }\right)$  becomes
\begin{align}
f\left(r_\text{min }+dr_\text{min }\right)=0, \label{eq4.12}
\end{align}
which shows that  $f_\text{min }+df_\text{min }=0$, so that   the charged particle does not change the minimum value. Therefore,  the weak cosmic censorship conjecture is valid  for the extremal black hole surrounded by quintessence dark energy. It is interesting to note that this result same with that in the extended phase space, the configuration of the black hole has not change after the absorption, the extremal black hole is still an extremal black hole.

For the near-extremal black hole, equation (\ref{eq4.2}) can not be used. With the condition $r_h=r_\text{min }+\epsilon $, we can expand equation (\ref{eq4.2}) at $r_\text{min }$, which leads to
\begin{align}
&dM=\frac{ Q}{r_{\min }}{dQ} \nonumber\\
&~~~~~+\left( Q^2{dr}_{\min }+\frac{3}{2} r_{\min } \left(\frac{2 r_{\min }{}^3}{l^2}-{ar}_{\min }{}^{-3 \omega } \omega  (1+3 \omega)\right){dr}_{\min }- Q r_{\min } {dQ}\right)r_{\min }{}^{-3}\epsilon \nonumber\\
&~~~~~+\frac{1}{4} r_{\min }{}^{-4-3 \omega} \left(4{  }Q r_{\min }{}^{1+3 \omega }{dQ} +{dr}_{\min } \left(-8 Q^2 r_{\min }{}^{3 \omega }+9 a r_{\min } \omega  \left(1+4 \omega +3 \omega ^2\right)\right)\right) \epsilon ^2 \nonumber\\
&~~~~~+\mathcal O(\epsilon^3). \label{eq4.13}
\end{align}
By combining  equations (\ref{eq4.10}) and (\ref{eq4.13}) we have
\begin{align}
df_\text{min }=\frac{1}{2}{  }r_{\min }{}^{-4-3 \omega } \left(4 r_{\min }{}^{1+3 \omega }-9 a \omega  \left(2 \omega +3 \omega ^2-1\right)\right) \epsilon ^2{dr}_{\min }  +\mathcal O(\epsilon^3). \label{eq4.14}
\end{align}
We do the same calculation that we did in the extended phase space. For the near-extremal black hole, we also can define $\mathcal F_N=\frac{f{\left(r_{\min }+\text{dr}_{\min }\right)}}{\epsilon ^2}$.  Hence, we can get the express of  $\mathcal F_N$  in the normal phase  as

\begin{align}
\mathcal F_N=\frac{4 r_{\min } {dr}_{\min }-2 Q^2+3 a r_{\min }{}^{-3 w} \omega  \left(r_{\min }+3 r_{\min } \omega -3 {dr}_{\min } (1+\omega ) (3 \omega -1)\right)}{2 r_{\min }{}^4}-\frac{3}{l^2},  \label{eq4.15}
\end{align}
From the equation (\ref{eq4.15}), it is not easily determine the  value of $\mathcal F_N$ is positive or negative.  Similar with the extended phase space, we still have  Figure 8  and  Figure 9.
\begin{figure}[htb]
\centering
\includegraphics[scale=1.0]{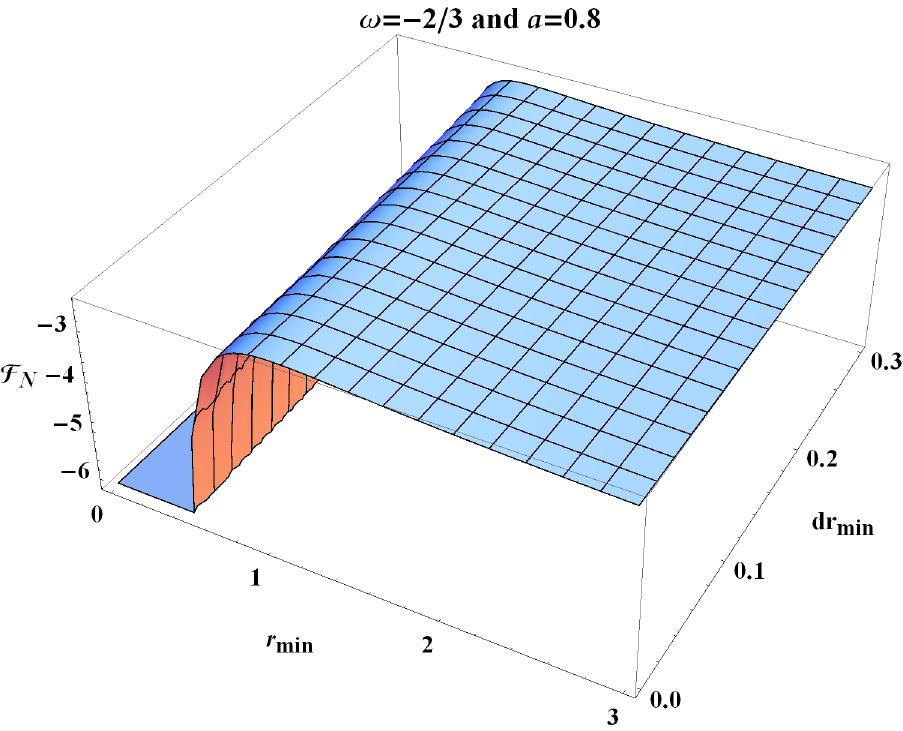}
\caption{The value of  $\mathcal F_N$ for $Q=0.5, l=a=1$ and $\omega=-0.5$.}
\label{fig:8}
\end{figure}
\begin{figure}[htb]
\centering
\includegraphics[scale=1.0]{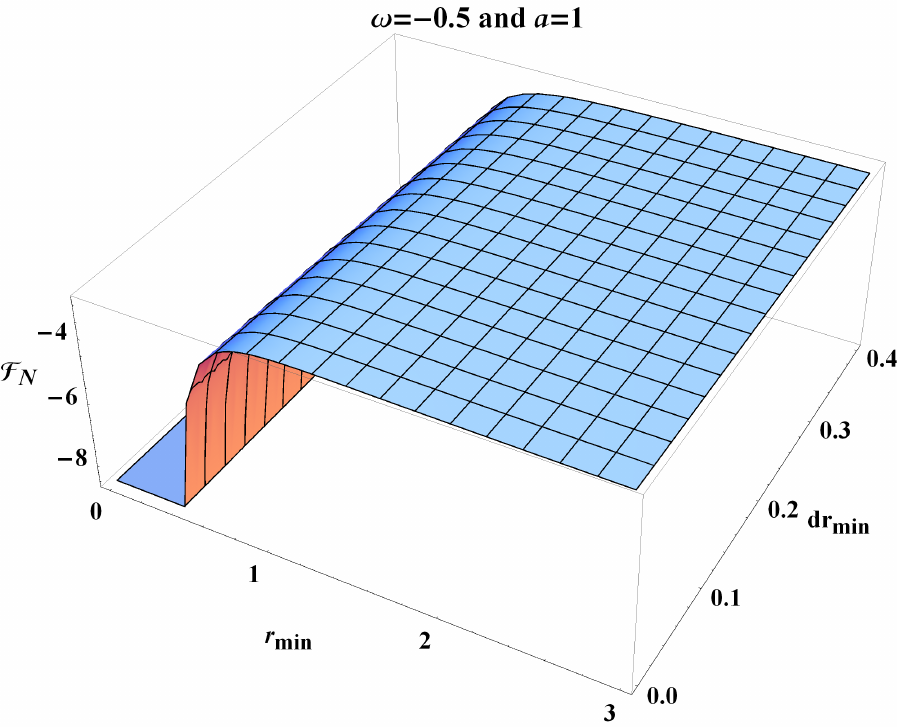}
\caption{The value of  $\mathcal F_N $ for $Q=0.5, l=a=1$ and  $\omega=-2/3$.}
\label{fig:9}
\end{figure}

In Figure 8 and Figure 9, there is no  region where the  value of $\mathcal F_N$ is positive. Concerning this result, it is important to note that the conclusion is very different from the extended phase space.
 In other word,  we always have  $ f\left(r_\text{min }+dr_\text{min }\right)<0$ for the near-extremal black hole in the normal phase space.  Therefore, the weak cosmic censorship conjecture of the near-extremal black hole surrounded by quintessence dark energy is valid under charged particle absorption in the normal phase space.

\section{Discussion and conclusions}\label{sec:5}

In the extended phase space, the cosmological parameter  has been set to be a dynamic variable and interpreted as the pressure. When the charged particle is absorbed by the black hole surrounded with quintessence dark energy,    the change of the energy and the charge of the  black hole were supposed to be equal to the conserved quantity of the charged particle. In this case, we found that the first law of thermodynamics was completely valid, but the second law of thermodynamics was violated for  extremal and near-extremal black holes. We also studied the validity of the weak cosmic censorship conjecture. We mainly  calculated the change of the minimum value of the function $f(M,Q,l,r)$ under the  charged particle absorption by studying the  minimum value of the function $f(M,Q,l,r)$.  Firstly,  we calculated the case of the extremal black holes.  We found that the  minimum value of the function was not changed under the charged particle absorption.  In other words, after the charge particle absorption, the extremal black hole stays extremal. Therefore, the weak cosmic censorship conjecture is valid for  the extremal black hole in the extended phase space.  However, our results shown that the  minimum value of the   function  under the absorption would appear the case of greater than zero. It has shown  that the near-extremal black hole could be overcharged, which was different from the extremal black hole.  That is, the cosmic censorship conjecture would be violated for the near-extremal black hole  surrounded by quintessence dark energy  after absorbs the charged particle in the extended phase space.

 In this paper, the results for the violation of  the weak cosmic censorship conjecture in the extended phase space are dissimilar from previous conclusions. In previous research \cite{Gwak:2017kkt}, they regarded the second order $\mathcal O(\epsilon^2)$ of $df_{\min}$ as a very small quantity, so that they  neglected  the contribution of $\mathcal O(\epsilon^2)$   to $df_{\min}$. Here,  we made precise comparisons and calculations at the minimum value of the function, and we obtained the   magnitudes relationship between $\delta$ and $\mathcal O(\epsilon^2)$. We found there was always a case of $f(r_{\min}+dr_{\min})>0$, which means that the weak cosmic censorship conjecture may be violated. And we still observes such violations under assuming $e<<1$.  Interestingly, when we taken $a=0$, the black hole returns  to the RN-AdS black hole, and the cosmic censorship conjecture was also invalid, which is different from the result in \cite{Gwak:2017kkt}.  

In normal phase space,  the cosmological constant is definite. We found that the first and second laws of thermodynamics  are both satisfied under the particle absorption. In addition, the weak cosmic censorship conjecture was also checked.  For the case of extremal black hole, we found that  the configuration of the black hole surrounded by quintessence dark energy  also has not  changed after absorbs charged particle, which is same as in the extended phase space. The result implies that the extremal black hole could not be overcharged in the normal phase space.  Interestingly, for the case of  near-extremal black hole, the minimum value of the function is still negative when the particle is absorbs into the black hole, which is different form that in the extended phase space. That is,  the weak cosmic censorship conjecture is valid for the near-extremal black holes in the normal phase space.

\section*{Acknowledgements}{This work is supported  by the National
Natural Science Foundation of China (Grant Nos. 11875095), and Basic Research Project of Science and Technology Committee of Chongqing (Grant No. cstc2018jcyjA2480).}
\bigskip

\bibliographystyle{unsrt}

\clearpage
\end{document}